\documentclass[12pt,english,preprint,superscriptaddress,showpacs,nofootinbib]{revtex4-1}
\usepackage[T1]{fontenc}
\usepackage[utf8x]{inputenc}
\usepackage{amsmath}
\usepackage{graphicx}
\usepackage{amssymb}
\usepackage{amsfonts}
\usepackage{epsfig}
\usepackage{subfigure}
\usepackage{graphicx}
\usepackage{pst-grad} 
\usepackage{pst-plot} 
\usepackage[labelsep = space,font={small}]{caption}
\usepackage{mathrsfs} 
\usepackage{float}

\usepackage{babel}
\addto\captionsenglish{}
\makeatletter

\@ifundefined{textcolor}{}
{%
 \definecolor{BLACK}{gray}{0}
 \definecolor{WHITE}{gray}{1}
 \definecolor{RED}{rgb}{1,0,0}
 \definecolor{GREEN}{rgb}{0,1,0}
 \definecolor{BLUE}{rgb}{0,0,1}
 \definecolor{CYAN}{cmyk}{1,0,0,0}
 \definecolor{MAGENTA}{cmyk}{0,1,0,0}
 \definecolor{YELLOW}{cmyk}{0,0,1,0}
 }

\makeatother

\begin{document}
\title {Magnetic Quantum Phases of Ultracold Dipolar Gases in an Optical Superlattice}
\thanks{X. Yin and L. Cao have contributed equally to this work.}
\author{Xiangguo Yin}
\email{xyin@physnet.uni-hamburg.de}
\affiliation{Zentrum f\"{u}r Optische Quantentechnologien, Universit\"{a}t Hamburg, Luruper Chaussee 149, D-22761 Hamburg, Germany}
\author{Lushuai Cao}
\email{lcao@physnet.uni-hamburg.de}
\affiliation{Zentrum f\"{u}r Optische Quantentechnologien, Universit\"{a}t Hamburg, Luruper Chaussee 149, D-22761 Hamburg, Germany}
\affiliation{The Hamburg Centre for Ultrafast Imaging,Luruper Chaussee 149,D-22761 Hamburg, Germany}
\author{Peter Schmelcher}
\email{pschmelc@physnet.uni-hamburg.de}
\affiliation{Zentrum f\"{u}r Optische Quantentechnologien, Universit\"{a}t Hamburg, Luruper Chaussee 149, D-22761 Hamburg, Germany}
\affiliation{The Hamburg Centre for Ultrafast Imaging,Luruper Chaussee 149,D-22761 Hamburg, Germany}

\begin{abstract}
 We propose an effective Ising spin chain constructed with dipolar quantum gases confined in a one-dimensional optical superlattice.
Mapping  the motional degrees of freedom of a single particle in the lattice onto a pseudo-spin results in
 effective transverse and longitudinal magnetic fields. This effective Ising spin chain exhibits a quantum phase transition from
 a paramagnetic to a single-kink phase as the dipolar interaction increases. Particularly in the single-kink phase,
 a magnetic kink arises in the effective spin chain and behaves as a quasi-particle in a pinning potential exerted by
 the longitudinal magnetic field.
 Being realizable with current experimental techniques, this effective Ising chain presents a unique platform for emulating
 the quantum phase transition as well as the magnetic kink effects in the Ising-spin chain
 and enriches the toolbox for quantum emulation of spin models by ultracold quantum gases.
\end{abstract}
\maketitle

\textit{Introduction.}---Ultracold quantum gases, equipped with an exceptionally good isolation from the environment and possessing an exquisite tunability of trap
and interaction parameters, have become a promising test bed for various fundamental physics problems,
ranging from cosmological \cite{sarkharov,blackhole} to condensed matter physics \cite{condensed_emulation}.
One paradigm is the emulation of quantum magnetism with ultracold quantum gases,
which has proved its flexibility by demonstrating, for instance, quantum magnetic
phase transitions
\cite{emulation_molecule_rey,emulation_rydberg,Ising-XY_sengstock,emulation_greiner,emulation_atom_nagerl,emulation_porbital},
magnetic frustration \cite{frustration_sengstock},
spin magnon states \cite{emulation_magnon}, as well as spin transport \cite{emulation_spintransport}.

One approach of spin chain emulations is through the Hamiltonian engineering of
dipolarly interacting optical lattice gases, such as
magnetic atoms \cite{emulation_magneticatom_tj}, polar molecules \cite{emulation_toolbox,emulation_molecule_rey,emulation_rey_dynamics}
and Rydberg atoms \cite{emulation_rydberg0,emulation_rydberg,emulation_rydberg2}.
In this approach, different internal states of the particles
are used to mimic the spin states, and an effective spin-spin interaction is 
induced by the dipolar interaction. This approach permits a flexible way of designing various spin models, such as 
Heisenberg and Ising models, accompanied by a rapid progress with respect to the experiment realizations
\cite{emulation_magneticatom_tj,molecular_exp}. An alternative
approach is based on a so-called pseudo-spin mapping applied to contact interacting lattice gases
\cite{pseudo_origin,emulation_greiner,emulation_atom_nagerl},
in which a pseudo-spin is mapped to the motional degrees of freedom of a single atom in the lattice.
This approach can benefit from the highly developed experimental manipulation of ultracold atoms in optical lattices,
and especially the pseudo-spin mapping allows an $in-situ$ spin state detection, enabled by the site-resolved particle parity measurement
\cite{insitu1,insitu2,emulation_greiner}.

In this letter we propose a new scheme for the quantum emulation of spin chains, which generates an effective Ising spin chain from a dipolar
superlattice gas by a pseudo-spin mapping, alternative to that in \cite{pseudo_origin,emulation_greiner,emulation_atom_nagerl}.
This scheme can combine advantages of the above mentioned two approaches, such as
the flexibility in the dipolar gas approach and the $in-situ$ spin detection ability in the pseudo-spin approach.
Particularly, this scheme supports a quantum emulation of magnetic kinks in Ising spin chains, which, to our best knowledge,
has not been thoroughly explored. Magnetic kinks are of central interest and at the heart of investigations on Ising chains,
and they are responsible for different dynamical processes,
such as the post-quench relaxation and thermal dynamics \cite{kink_thermal1,kink_thermal2,kink_relaxation}
as well as the Kibble-Zurek transition \cite{kink_KZM,kink_interference}.
The design and control of kinks in Ising chains can shed light on various stationary and dynamical effects of the spin system.
One reason for the limited access to these aspects is that in normal Ising chains, such as the transverse
Ising chain, the kink states lie only in the excited state spectrum and cannot be detected in the ground state. The effective Ising chain proposed
in this letter, however, presents a kink-dominated ground state, and
this allows for a direct access to various effects related to the magnetic kink,
which makes the effective Ising chain a unique platform for the quantum emulation of magnetic kink properties.

\begin{figure}[tbp]
\centering
\includegraphics[width=8.6cm]{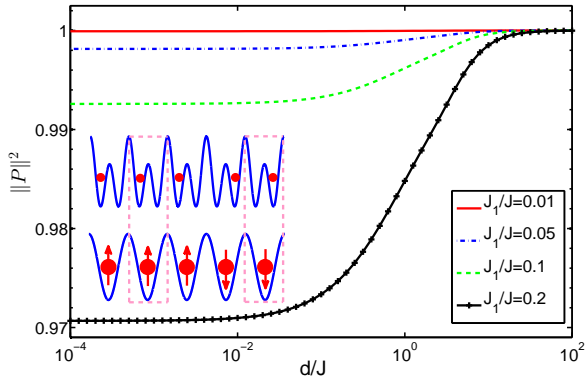}
\caption{(Color online). The spin-basis projection norm $\|P\|^2$ as a function of the scaled dipolar interaction strength $d/J$ for a five-supercell chain, for different hopping
strengths $(J,J_1)$. A sketch of the pseudo-spin mapping in this five-cell chain, is shown in the inset,
where the dashed boxes illustrate the pseudo-spin mapping.
}
\label{fig:Fig1}
\end{figure}

\textit{Spin model construction}.---We demonstrate in the following the construction of the dipolar-superlattice-gas (DSG) Ising chain
using bosonic dipolar gases, 
while the same procedure can be directly applied to fermionic gases. The dipolar gas considered here is composed 
of field aligned repulsively interacting dipoles,
and the 1D superlattice is chosen as a double-well 
superlattice, a widely used test bed for various quantum phenomena \cite{dw_exp1,dw_entangle,dw_topological},
in which each supercell contains two sites and mimics a double well, as sketched in the inset of Fig. \ref{fig:Fig1}.
We focus on the special filling of one particle per supercell,
the generalization to lower fillings being straightforward.
The dipolar gas confined in a double-well superlattice of $N$ supercells
($i.e.$ $2N$ sites) is described by the extended Bose-Hubbard Hamiltonian
\cite{dipolar_EBH,dipolar_review1,dipolar_review2}, which reads:
\begin{eqnarray}
\nonumber\hat{H}_{BH}&=&-J\sum_{j=1}^{N}(\hat{a}_{2j}^{\dagger}\hat{a}_{2j-1}+H.C.)
-J_{1}\sum_{j=1}^{N-1}(\hat{a}_{2j}^{\dagger}\hat{a}_{2j+1}+H.C.)\\
&&+\sum_{i<j\in[1,2N]}V_d(j-i)\hat{n}_{i}\hat{n}_{j}+U\sum_{i=1}^{2N}\hat{n}_i(\hat{n}_i-1)/2.  \label{H1}
\end{eqnarray}
In this Hamiltonian $\hat{a}_{2j/2j-1}^{\dagger}$ ($\hat{a}_{2j/2j-1}$) creates (annihilates) a boson in the right/left 
site of the $j$-th supercell, with $\hat{n}_i\equiv\hat{a}_{i}^{\dagger}\hat{a}_{i}$ denoting the density operator of the $i$-th site.
The first two terms of $\hat{H}_{BH}$ are the intra- and inter-supercell hopping terms, respectively.
The off-site dipolar interaction is described by
the third term of $\hat{H}_{BH}$ and, to a good approximation, is defined as $V_d(j-i)=d\lambda^{-3}|j-i|^{-3}$
with $d$ denoting the dipolar interaction strength and the neighboring sites distance $\lambda$ being normalized to unity.
The on-site interaction, induced by both dipolar and contact interactions, is described by the last term of $\hat{H}_{BH}$.
We assume a hard-core superlattice gas with a large $U$. We note, however, that our main results are not restricted to this assumption.
A common condition $J\gg J_1$ in double-well superlattices is applied here, and the interplay between the dipolar interaction and two tunneling processes
determines the ground state of $\hat{H}_{BH}$, on top of which the DSG Ising chain is constructed.

Dipolar gases in optical lattices of different geometries present various nontrivial phases 
\cite{dipolar_review1,dipolar_review2,dipolar_phase,dipolar_phase2,dipolar_phase3}, and loaded in a
double-well superlattice, the dipolar gases with a filling of one particle per supercell will reside in a Mott-like ground state 
under the condition of $J\gg J_1\approx0$ and $d\gg J_1$. In the Mott-like ground state, each supercell only accommodates a single
particle, and while confined in a supercell, each particle still possesses a finite mobility through the hopping between the two sites
of the host supercell.
To construct the effective Ising spin chain on top of the Mott-like ground state, we apply a pseudo-spin mapping
and regard each particle as a pseudo-spin, mapping the state of the particle occupying the left/right site in the supercell to the
$\uparrow$/$\downarrow$ pseudo-spin state.
The dipolar superlattice gas in the Mott-like state is then transferred to a pseudo-spin chain, and correspondingly the Hilbert space of the dipolar superlattice gas
is truncated to a subspace spanned by the so-called spin-chain states
$\{|\vec \alpha\rangle=\prod_{i=1}^{N}|\alpha_i\rangle_i\}$, 
with $\alpha_i\in\{\uparrow,\downarrow\}$ denoting the $i$-th pseudo-spin. Under this pseudo-spin mapping and Hilbert space truncation, $\hat{H}_{BH}$ is transferred 
to an Ising-type spin chain Hamiltonian:
\begin{eqnarray}
\hat{H}_{Ising}&=&-J\sum_{i=1}^{N}\hat{\sigma}_{x,i}-\sum_{i<j=1}^{N}W_d(j-i)\hat{\sigma}_{z,i}\hat{\sigma}_{z,j}\\\nonumber
&&+\sum_{i=1}^{N}U_d(i)\hat{\sigma}_{z,j}+E_0, \label{H2}
\end{eqnarray}
in which $\hat\sigma_{z,i}\equiv\hat{n}_{2i-1}-\hat{n}_{2i}$ and $\hat\sigma_{x,i}\equiv\hat{a}^\dagger_{2i-1}\hat{a}_{2i}+H.c.$ are the Pauli
matrices of the $i$-th pseudo-spin with the quantization axis chosen as the $z$-axis. In $\hat{H}_{Ising}$ the first term describes an 
effective transverse magnetic field, which originates from the intra-supercell hopping. The second and third terms of $\hat{H}_{Ising}$
refers to an Ising-type spin-spin interaction and a longitudinal magnetic field, respectively, which are derived from the dipolar interaction, with 
$W_d(l)=[V_{d}\left(2l-1\right)+V_{d}\left(2l+1\right)-2V_{d}\left(2l\right)]/4$
and $U_d(i)=[V_d(2N-2i+1)-V_d(2i-1)]/4$. The longitudinal magnetic field exhibits a spatial dependence, and it polarizes oppositely along
the $z$- and $(-z)$-axis directions on the left and right halves of the chain, respectively, and the magnitude decays
from the edges to the center of the chain with a power law.
This longitudinal magnetic field manifests itself as a key term which deviates $\hat{H}_{Ising}$ from that of a transverse Ising chain, and is
responsible for the emergence of the magnetic kinks in the chain. Finally
a constant term $E_0=\sum_{l=1}^{2N-1}V_{d}\left(l\right)\left(2N-l\right)/4-V_d(1)N/4$
also arises from the dipolar interaction. 

As $W_d(l)$ and $U_d(l)$ decay with a power law, for not too strong $d$
it is a good approximation to just keep in $\hat{H}_{Ising}$ the terms corresponding to $U_d(1)$, $U_d(N)$ and $W_d(1)$,
which are the longitudinal magnetic fields on the left/right edge sites and the nearest-neighbor interactions, respectively.
We denote the Hamiltonian in this approximation as $\hat{H}_{Ising}^{NN}$.
In the following we will investigate and compare the quantum phase transition (QPT) and also magnetic kink properties obtained from $\hat{H}_{Ising}$
and $\hat{H}_{Ising}^{NN}$.

We numerically verify the validity of the pseudo-spin mapping, by calculating the spin-basis projection norm 
$\|P\|^2\equiv\sum_{\vec \alpha}|\langle\vec\alpha|G\rangle|^2$ in a five-cell superlattice, in which $|G\rangle$ is the ground state of
$\hat{H}_{BH}$
and the summation over $\vec \alpha$ runs through all the spin-chain states. Fig. \ref{fig:Fig1} shows that $\|P\|^2$ approaches
unity as $d$ dominates over $J_1$. This confirms that in the related interaction regime the truncation of
$|G\rangle$ to the spin-chain basis is a good approximation and the pseudo-spin mapping is valid.
Our proceeding investigations will then base on the parameter set
$(J,J_1)=(1,0.01)$, with varying $d$, which relates to the red curve in Fig. \ref{fig:Fig1}.
The pseudo-spin mapping of the five-cell superlattice is sketched in the inset of Fig. \ref{fig:Fig1}, where the 
spin chain configuration refers to a kink state.

\textit{Quantum Phase transition}.---The DSG Ising chain exhibits a QPT driven by the dipolar interaction.
In the weak interaction regime, it resides in a paramagnetic
phase polarized by the transverse magnetic field. In the strong interaction regime, the system enters a single-kink phase,
instead of a ferromagnetic phase occurring in a transverse Ising chain. In the single-kink phase, the spin-spin interaction energetically
prefers all spins to be polarized in the same direction, while the longitudinal field tends to align the spins on the left half to the opposite
direction than on the right half of the chain.
The interplay between these two dominant mechanisms in the single-kink phase induces two ferromagnetic
domains in the chain, with opposite polarization in the $z$-axis and $(-z)$-axis directions of the left and right domains, respectively.
The boundary between the two domains forms a kink, and the position of the boundary is not fixed for the case of a relatively weak longitudinal field,
which leads to a finite mobility of the kink.
In this section we provide a characterization of the QPT,
and the quasi-particle behavior of the kink will be explored in the following section.

\begin{figure}[tbp]
\centering
\includegraphics[width=8.6cm]{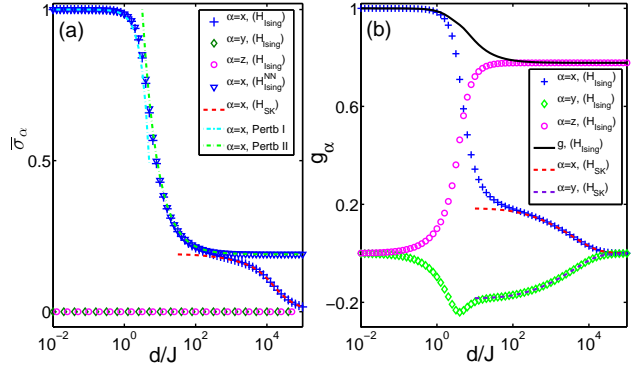}
\caption{(Color online). (a) The magnetization $\overline{\sigma}_{\alpha}$ and (b) the NN spin-spin correlations $g_{\alpha}$ of a ten-spin DSG Ising chain as a function of the scaled dipolar interaction strength $d/J$.
In both subfigures, the $+$/$\diamond$/$\circ$ curves represent the x/y/z components of the corresponding quantities obtained through exact diagonalization (ED) with
$\hat{H}_{Ising}$,
and the red/brown dashed curves are the x/y components of the corresponding quantities obtained through ED with the reduced Hamiltonian $\hat{H}_{SK}$.
In (a), $\overline{\sigma}_x$ is also calculated using $\hat{H}^{NN}_{Ising}$, by ED ($\bigtriangledown$), as well as analytically by
perturbation theory: Pertb I (cyan dash-dot) and Pertb II (green dash-dot) \cite{supp}.}
\label{fig:Fig2}
\end{figure}

The QPT in the DSG Ising chain can be characterized by the spin chain magnetization
$\overline{\boldsymbol{\sigma}}=(\overline{\sigma}_x,\overline{\sigma}_y,
\overline{\sigma}_z)$, with $\overline{\sigma}_\alpha\equiv\sum_{i=1}^{N}\langle\hat\sigma_{\alpha,i}\rangle/N$.
We calculate $\overline{\boldsymbol{\sigma}}$ for a ten-site
chain both analytically and numerically, by perturbation theory \cite{supp} and exact diagonalization, respectively, with a good coincidence
between the results as shown in Fig. \ref{fig:Fig2}(a), where the results based on $\hat{H}_{Ising}$ and $\hat{H}^{NN}_{Ising}$ are compared.
Fig. \ref{fig:Fig2}(a) shows that the magnetization in the whole interaction
regime obeys $\overline{\sigma}_y=\overline{\sigma}_z=0$,
and $\overline{\sigma}_x$ exhibits two plateaus, which evidences a QPT.
On the first plateau of $\overline{\sigma}_x$ in the weak interaction regime, the system resides in
a paramagnetic phase, and is polarized by the transverse field. On the second plateau in the strong interaction regime, the system enters the 
single-kink phase. In this phase, as will be discussed in the following section, the slow decay of $\overline{\sigma}_x$
on the second plateau obtained from $\hat{H}_{Ising}$ indicates a reduction of the kink mobility as the interaction increases.
Compared to $\hat{H}_{Ising}$, $\hat{H}_{Ising}^{NN}$ can very well reproduce the QPT from the paramagnetic to the single-kink phase,
while it fails to characterize the reduction of the kink mobility for sufficiently strong interactions.
This deviation between the results of $\hat{H}_{Ising}$ and $\hat{H}_{Ising}^{NN}$ reflects the fact that the
longitudinal field on the non-edge sites neglected in $\hat{H}_{Ising}^{NN}$ dominates the mobility reduction of the kink. 

Furthermore, in Fig. \ref{fig:Fig2}(b) we also calculate the nearest-neighbor (NN) spin-spin correlation 
$g\equiv\sum_{\alpha=x,y,z}g_{\alpha}$
to illustrate the QPT, where $g_{\alpha}=\frac{1}{N-1}\sum_{i=1}^{N-1}
\langle\hat\sigma_{\alpha,i}\hat\sigma_{\alpha,i+1}\rangle$. In the paramagnetic phase, only $g_{x}$
takes a non-zero value, and the overall spin-spin correlation $g$
coincides with $g_{x}$. In the single-kink phase,
despite a net zero magnetization in $z$-direction, the two ferromagnetic domains induce a finite value for $g_{z}$,
and $g$ becomes dominated by $g_{z}$.
In the single-kink phase, $g_{x}$ and $g_{y}$ 
also present a moderate decay, which is induced again by the mobility reduction of the kink.

\textit{Kink-state phase}.---We now turn to a detailed investigation of the kink properties in the single-kink phase. In this phase, the single-kink states, which are
defined as $|n\rangle_{K}=\prod^{n}_{i=1}|\uparrow\rangle_i\otimes\prod^{N}_{i=n+1}|\downarrow\rangle_i$ $(n\in[1,N-1])$, are
energetically decoupled from other spin states, and they dominate the lowest band in the energy spectrum.
Within the single-kink basis $\{|n\rangle_{K}\}|_{n=1}^{N-1}$, one can derive a reduced Hamiltonian from $\hat{H}_{Ising}$, which reads:
\begin{equation}
 \hat{H}_{SK}=-\sum_{n=1}^{N-2}J(\alpha^\dagger_{n+1}\alpha_n+H.c.)+\sum_{n=1}^{N-1}\epsilon_n\alpha^\dagger_n\alpha_n,
\end{equation}
where $\alpha^\dagger_n$ ($\alpha_n$) creates (annihilates) a kink between the $n$ and $n+1$ spins. $\hat{H}_{SK}$ illustrates that the kink
behaves as a quasi-particle hopping along the spin chain, in which the NN hoppings are induced by the transverse magnetic field. The non-edge
longitudinal magnetic field originated from the dipolar interaction generates a pinning trap to the kink, which is described by
the second term of $\hat{H}_{SK}$ with $\epsilon_n\equiv{}_K\langle n|\hat{H}_{Ising}|n\rangle_K$ \cite{HSK}. Moreover,
in the weak interaction regime within the single-kink phase, the NN hoppings are dominant over the longitudinal-field induced pinning trap,
and the kink can move freely along the chain, resembling a free particle in a box potential. In the deep regime of the single-kink phase,
however, the pinning trap becomes dominant and gradually pins the kink to the middle of the chain, which effectively leads to a vanishing
mobility of the kink.

Within the single-kink basis, one obtains $\hat\sigma_{x,i}=\alpha^\dagger_{i-1}\alpha_i+H.c.$ as well as
$\hat\sigma_{x,i}\hat\sigma_{x,i+1}=\alpha^\dagger_{i-1}\alpha_{i+1}+H.c.$, and 
$\overline{\sigma}_x$ and $g_x$ turn to measure the NN and
next-nearest-neighbor (NNN) spatial correlations of the kink in the chain, respectively. Then the mobility reduction of the kink
leads to the vanishing of the NN and NNN correlations and consequently the decay of 
$\overline{\sigma}_x$ and $g_x$ in the strong interaction regime.
The same analysis also holds for $g_y$, which turns out to be
$g_y=-g_x$.
In Fig. \ref{fig:Fig2}, we calculate the magnetization and spin-spin correlation from $\hat{H}_{SK}$, 
which reproduces very well the numerical results by $\hat{H}_{Ising}$ in the strong interaction regime.

\begin{figure}[tbp]
\centering
\includegraphics[width=8.6cm]{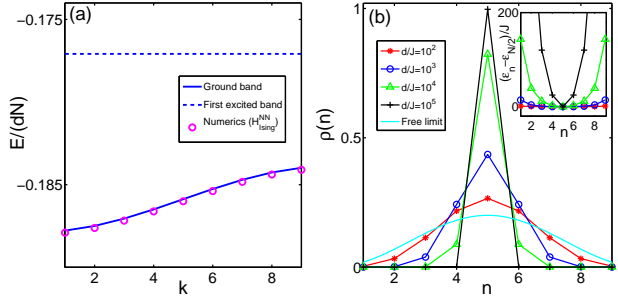}
\caption{(Color online). (a) The energy spectrum of a ten-spin DSG Ising chain in the single-kink phase, calculated by exact diagonalization (circles)
and perturbation theory (solid line). The band of the doubly degenerate eigenstates lying nearest to the ground band is also shown with a dashed line.
The numerical results are obtained for $d/J=100$.
(b) The spatial probability of the kink with different interaction strengths in comparison with the ideal free-particle limit.
For each probability curve in figure (b), the corresponding pinning potential substracted by the potential minimum
is shown in the inset with the same line type.}
\label{fig:Fig3}
\end{figure}

Now we explore the quasi-particle behavior of the kink in terms of the energetic band structure and the mobility of the kink, for which
analytical and numerical calculations are performed in a ten-spin chain.
Fig. \ref{fig:Fig3}(a) presents the energy spectrum of the spin chain near the onset of the 
single-kink phase, with a focus on the lowest band.
The degenerate perturbation calculation based on $\hat{H}_{Ising}^{NN}$ gives that the eigenstates in the ground band are 
$\left\vert \psi\right\rangle _{k}=\sum_{n=1}^{N-1}\sqrt{2/N}\sin\left(kn\pi/N\right) \left\vert n\right\rangle _{K}$,
with corresponding eigenenergies $E_{k}=-(N-3)W_{d}(1)+2U_{d}(1)-2J\cos\left( k\pi/N \right)$ ($k\in[1,N-1]$), which illustrates the quasi-particle behavior
of the kink \cite{kink_classical}.
Furthermore, Fig. \ref{fig:Fig3}(a) also shows the eigenstates lying above the ground band, which are the doubly degenerate ferromagnetic states
$\prod^{N}_{i=1}|\uparrow\rangle_i$ and $\prod^{N}_{i=1}|\downarrow\rangle_i$. 
The mobility of the kink can be characterized by $\rho(n)=|{}_{K}\langle n|G\rangle_I|^2$ 
($|G\rangle_I$ denotes the ground state of $\hat{H}_{Ising}$), which defines the probability of finding the kink between the $n$ and $n+1$ spins.
Having verified $\sum_{n=1}^{N-1}\rho(n)$ approaching unity, which guarantees the validity of the truncation of the ground
state to the kink basis,
we observe in Fig. \ref{fig:Fig3}(b) that as the dipolar interaction increases, the pinning trap grows stronger (shown in the inset) and
the kink indeed becomes more and more localized to the middle of the chain, $i.e.$ the pinning trap bottom,
which reflects the mobility reductions. 

It is also worth mentioning that in the single-kink phase,
the strong dipolar interaction guarantees that not only the ground state but all the eigenstates in the ground band of $\hat{H}_{Ising}$
are also the lowest eigenstates of $\hat{H}_{BH}$ describing the original dipolar superlattice gas. Then one can extend the
ground-state QPT to kink-related dynamical processes in this phase.

\textit{Experimental implementation}.---In the following we briefly discuss the necessary condition for the experimental
implementation of the DSG Ising chain.
The DSG Ising chain possesses a major advantage for its experimental implementation, $i.e.$, the effective spin chain is constructed on top of
the true ground state of the dipolar superlattice gas, and the experimental realization does not require additional preparations other than
relaxing the system to its ground state. Moreover, both bosonic and fermionic dipolar gases can be used for the realization.
This allows for various dipolar gases as a candidate for the experimental realization of the DSG Ising chain, such as magnetic atoms, 
polar molecules, Rydberg atoms, as well as the recently proposed laser dressed alkaline-earth atoms \cite{alkalineatom1,alkalineatom2}.
Then a general requirement is the flexible tunability of the ratio $d/J$.
Such a tunability can be achieved by either tuning the dipolar interaction strength or, more simply, tuning the superlattice barriers.

The generation and direct detection of magnetic kinks still remains a challenge in condensed matter physics, while the DSG Ising chain with the 
pseudo-spin mapping holds the possibilities to utilize the $in-situ$ imaging technique developed for the optical lattices for a direct and
global detection of the kinks. Besides, quantum interference based measurements have also been proposed for the detection of kinks in the Ising chain
\cite{kink_interference}.

\textit{Outlook}.---We have demonstrated the construction of an effective Ising chain with dipolar superlattice gases,
which holds the promise for the quantum emulation of 
QPTs and magnetic kink states. This DSG Ising chain construction is not restricted to a particular filling in a plain superlattice
without additional external trap. A lower filling and an external trap, $e.g.$ a dipole trap, will induce the hopping of the spins and a modified
longitudinal field, respectively, which can enhance the emulation flexibility of the DSG Ising chain.
Such DSG Ising chains can also be generalized to high-spin systems, by increasing the sites in each supercell.
Given the flexibilities and robustness of the DSG Ising chains, we expect that it can add new elements to the toolbox
of the emulation of quantum magnetism with ultracold quantum gases.

\begin{acknowledgments}
 The authors gratefully acknowledge funding by the Deutsche Forschungsgemeinschaft in the framework of the SFB 925 ``Light induced
dynamics and control of correlated quantum systems''.
\end{acknowledgments}

\bibliographystyle{apsrev4-1}
\bibliography{references}

\newpage
\textbf{Supplemental Material}
\newline

We perform perturbation thoery with respect to $\hat{H}_{Ising}^{NN}=\hat{H}_{x}+\hat{H}_{i}+\hat{H}_{z}$, 
in which we define the transverse field, spin-spin interaction, and longitudinal field term as
\begin{eqnarray}\nonumber
\hat{H}_{x}&=&-J\sum_{i=1}^{N}\hat{\sigma}_{x,i},\\\nonumber
\hat{H}_{i}&=&-\sum_{i=1}^{N-1}W_d(1)\hat{\sigma}_{z,i}\hat{\sigma}_{z,i+1},\\\nonumber
\hat{H}_{z}&=&U_d(1)(\hat{\sigma}_{z,1}-\hat{\sigma}_{z,N}). 
\end{eqnarray}

\section{Perturbation thoery in the paramagnetic phase (Pertb I)}
For $d/J\ll 1$, the transverse magnetic field $\hat{H}_x$ dominates, and the
spin-spin interaction $\hat{H}_i$ and longitudinal magnetic field $\hat{H}_z$ can be considered as perturbations.

In the zeroth order perturbation, the ground state, $i.e.$ the ground state of $\hat{H}_x$, is given by
$\left\vert0\right\rangle _{J}=\prod^{N}_{i=1}\left\vert \rightarrow \right\rangle _{i}$, with  
$\left\vert \rightarrow\right\rangle_{i} \equiv \frac{1}{\sqrt{2}}\left(\left\vert\uparrow\right\rangle_{i}+
\left\vert\downarrow\right\rangle_{i}\right)$ and $\left\vert \leftarrow\right\rangle_{i}
\equiv \frac{1}{\sqrt{2}}\left(\left\vert\uparrow\right\rangle_{i}-\left\vert\downarrow\right\rangle_{i}\right)$.
The corresponding ground eigenenergy is $E_{0}^{(0)}=-JN$.
The first exctied band is composed of $N$-fold degenerate excited eigenstates, which are generated by flipping one spin
in $\left\vert0\right\rangle _{J}$ from $\left\vert\rightarrow\right\rangle$ to $\left\vert\leftarrow\right\rangle$.
These eigenstates are written as $\left\vert i\right\rangle_{J,1}=\hat\sigma_{z,i}\left\vert0\right\rangle_J$
with the degenerate eigenenergy $E_{1}^{(0)}=-J(N-2)$. The second excited band is composed of $\frac{N(N-1)}{2}$-fold degenerate eigenstates,
corresponding to flipping two spins in $\left\vert0\right\rangle _{J}$. 
The eigenenergy of the degenerate second-excited band is $E_{2}^{(0)}=-J\left( N-4\right)$.

In the first order perturbation the states
$\left\vert1\right\rangle_{J,1}$ and $\left\vert N\right\rangle_{J,1}$ in the first excited band and
$\left\vert i,i+1\right\rangle _{J,2}=\hat\sigma_{z,i}\hat\sigma_{z,i+1}
\left\vert0\right\rangle _{J}$ in the second excited band are coupled to $\left\vert0\right\rangle _{J}$,
and the ground state is obtained as
\begin{eqnarray}\nonumber
\left\vert \psi^{(1)} _{0}\right\rangle =\left\vert 0\right\rangle
_{J}-\sum_{i=0}^{N}\frac{W_d(1)}{E_{0}^{(0)}-E_{2}^{(0)}}\left\vert
i,i+1\right\rangle _{J,2
}+\frac{U_d(1)}{E_{0}^{(0)}-E_{1}^{(0)}}(\left\vert
1\right\rangle _{J,1}-\left\vert
N\right\rangle _{J,1}).
\end{eqnarray}
We also apply a normalization to $\left\vert \psi^{(1)} _{0}\right\rangle$, which becomes
\begin{eqnarray}\nonumber
\left\vert \psi^{(1)} _{0}\right\rangle =A_{J}\left[ \left\vert 0\right\rangle
_{J}-\frac{u}{2}\left( \left\vert N\right\rangle _{J,1}-\left\vert
1\right\rangle _{J,1}\right) +\frac{v}{4}\sum_{i=1}^{N-1}\left\vert
i,i+1\right\rangle _{J,2}\right],
\end{eqnarray}
with $u=-U_{d}\left( 1\right) /J$, $v=W_{d}\left( 1\right) /J$,
and the normalization coefficient \\
$A_{J}=1/\sqrt{1+u^{2}/2+\left( N-1\right) v^{2}/16}$. 
The eigenenergy of the ground state is obtained as:
\begin{eqnarray}\nonumber
E^{(1)}_{0}=E_{0}^{(0)}+{}_{J}\left\langle 0\right\vert \left( \hat{H}%
_{i}+\hat{H}_{z}\right) \left\vert 0\right\rangle _{J}=E_{0}^{(0)}.
\end{eqnarray} 
The first order perturbation treatment lift the degeneracy of the first excited state band, and the eigenstates become:
\begin{eqnarray}\nonumber
\left\vert \psi _{n}^{(1)}\right\rangle
_{J}=\sum_{i=1}^{N}C_{i}^{n}\left\vert i\right\rangle _{J,1},
\end{eqnarray} 
with $C_{i}^{J}=\sqrt{2/\left( N+1\right) }\sin \left( ki\pi /\left(N+1\right) \right)$. Correspondingly, the eigenenergies
are derived as:
\begin{eqnarray}\nonumber
E_{k}^{(1)}=-J\left( N-2\right)-2W_{d}\left( 1\right) \cos \left( k\pi /N\right).
\end{eqnarray}

In the first order perturbation approximation, the expectation values of the magnetization and nearest-neighbor spin-spin correlations
have the analytic form:
\begin{eqnarray}\nonumber
\overline{\sigma}_x &=&A_{J}^{2}\left\{ 1+\frac{1}{2N}\left[u^{2}\left( N-2\right) +\frac{v^{2}}{8}\left( N-4\right) \left( N-1\right)\right] \right\},\\\nonumber
g_{NN}^{x} &=&A_{J}^{2}\left\{ 1+\frac{1}{2\left( N-1\right) }\left[ u^{2}\left( N-3\right) +\frac{v^{2}}{8}\left(N^{2}-6N+9\right) \right] \right\},\\\nonumber
g_{NN}^{z} &=&\frac{v}{2}.
\end{eqnarray}

\section{ Perturbation thoery in the single-kink phase (Pertb II)}

For $d/J\gg 1$, $\hat{H}_z$ and $\hat{H}_i$
dominate, and $\hat{H}_x$ is considered as a perturbation. 

In the zeroth order perturbation, the single-kink states
$|n\rangle_{K}=\prod^{n}_{i=1}|\uparrow\rangle_i\otimes\prod^{N}_{i=n+1}|\downarrow\rangle_i$ $(n\in[1,N-1])$
are the $(N-1)$-fold degenerate ground states, with degenerate eigenenergy 
\begin{eqnarray}\nonumber
E_{0}^{(0)}=-(N-3)W_{d}\left( 1\right) +2U_{d}\left( 1\right).
\end{eqnarray}
The first order perturbation treatment leads to the coupling within the single-kink basis and lifts the degeneracy.
The second order perturbation treatment then couples the single-kink states to other spin states. The spin states
$\left\vert n,i\right\rangle=\hat\sigma_{x,i}|n\rangle_{K}$
($i\in[1,n)\bigcup(n+1,N]$), which correpsond to flip a spin beyond the kink boundary sites,
are coupled to the single-kink states in the second order perturbation treatment. States
$\left\vert \Uparrow\right\rangle$ and $\left\vert \Downarrow\right\rangle$, which correpsond to ferromagnetic states polarizing to
$z$- and $-z$-axis directions, respectively, also couple to the single-kink states in the second perturbation treatment.

The first order degenerate perturbation approximation applies diagonalization of $\hat{H}_x$ with respect to the single-kink basis, 
and the obtained eigenstates in the ground band are written as:
\begin{eqnarray}\nonumber
 \left\vert
\psi ^{(1)}_{k}\right\rangle =\sum_{i=1}^{N-1}C^k_{i}\left\vert i\right\rangle
_{K},
\end{eqnarray}
with $C^k_{i}=\sqrt{2/N}\sin \left( ki\pi /N\right)$. The corresponding eigenenergies are
\begin{eqnarray}\nonumber
E^{(1)}_{k}=-(N-3)W_{d}\left( 1\right) +2U_{d}\left( 1\right) -2J\cos \left( k\pi/N\right).
\end{eqnarray}%

The ground state in the second order degenerate perturbation approximation is derived by the formula:
\begin{equation}\nonumber
\left\vert \psi^{(2)}_{k} \right\rangle =\left\vert \psi_{k}^{(1)}\right\rangle
+\sum_{\alpha}\frac{\left\langle \alpha\right\vert \hat{H}_{x}\left\vert \psi
^{(1)}_{k}\right\rangle}{E_{0}^{(0)}-E_{\alpha}^{(0)}}(\left\vert
\alpha\right\rangle +\sum_{p\neq k}\frac{\left\langle\psi^{(1)}_p\right\vert
\hat{H}_{x}\left\vert \alpha\right\rangle }{E^{(1)}_{k}-E^{(1)}_{p}}%
\left\vert \psi^{(1)}_p\right\rangle),
\end{equation}%
where $\left\vert \alpha\right\rangle$ denotes the spin states beyond the single-kink basis.
The final expression of $\left\vert \psi^{(2)}_{k}\right\rangle$ is given by:
\begin{eqnarray}\nonumber
\nonumber
\left\vert \psi^{(2)}_k \right\rangle &=&\left\vert \psi ^{(1)}_k\right\rangle
+A_{0}\left( C_{1}\left\vert \Uparrow\right\rangle+C_{N-1}\left\vert
\Downarrow\right\rangle\right) 
+A_{2}\left[ \sum_{n=2}^{N-1}C_{n}\left\vert
n,1\right\rangle +\sum_{n=1}^{N-2}C_{n}\left\vert n,N\right\rangle \right]  \\\nonumber
&&+A_{3}\left[\sum_{n=1}^{N-1}C_{n}\left( \sum_{i\in[2,n-2]}+\sum_{i\in[n+3,N-1]}\right)\left\vert
n,i\right\rangle+
\sum_{n=1}^{N-3}\left( C_{n}+C_{n+2}\right)\left\vert n,n+2\right\rangle\right],
\end{eqnarray}%
in which $A_{0}=\frac{J}{2\left( -U_{d}\left( 1\right) -W_{d}\left( 1\right)
\right) }$, $A_{2}=\frac{J}{2\left( -U_{d}\left( 1\right) +W_{d}\left(
1\right) \right) }$, $A_{3}=\frac{J}{4W_{d}\left( 1\right) }$.

The corresponding expectation value of magnetization and NN spin-spin correlations in the 
second order perturbation theory are then obtained:
\begin{eqnarray}\nonumber
\overline{\sigma}_x &=&\frac{2}{N}\left[ \cos (\pi/N)+A_{3}\left( N-4\right) +2A_{2}+2\left( A_{0}-A_{2}+A_{3}\right)
C_{1}^{2}+2A_{3}\sum_{n=1}^{N-3}C_{n}C_{n+2}\right],\\\nonumber
g_{NN}^x&=&\frac{1}{\left(N-1\right)}\left[2\sum_{n=1}^{N-3}C_{n}C_{n+2}+4A_{3}cos(\pi/N)+4(A_{0}+A_{2}-A_{3})C_{1}C_{2}+4A_{3}\sum_{n=1}^{N-4}C_{n}C_{n+3}\right],\\ \nonumber
g_{NN}^z&=&\frac{1}{\left(N-1\right)}\{N-3+A_{3}^{2}\left(N-7\right)\left(N-4+2\sum_{n=1}^{N-3}C_{n}C_{n+2}\right)+2A_{2}^{2}\left(N-5\right) \\\nonumber
&&+2C_{1}^{2}\left[A_{0}^{2}\left(N-1\right)-A_{2}^{2}\left(N-5\right)+A_{3}^{2}\left(N-7\right)\right]\}.
\end{eqnarray}
\end{document}